# Direct observation of local Mn-Mn distances in the paramagnetic compound $CsMn_xMg_{1-x}Br_3$

A. Furrer[1], Th. Strässle[1], J. P. Embs[1], F. Juranyi[1], V. Pomjakushin[1], M. Schneider[1] and K. W. Krämer[2]

[1]Laboratory for Neutron Scattering, Paul Scherrer Institute, CH-5232 Villigen PSI, Switzerland

[2]Department of Chemistry and Biochemistry, University of Bern, CH-3012 Bern, Switzerland

**Abstract:**

We introduce a novel method for local structure determination with a spatial resolution of the order of 0.01 Å. It can be applied to materials containing clusters of exchange-coupled magnetic atoms. We use neutron spectroscopy to probe the energies of the cluster excitations which are determined by the interatomic coupling strength J. Since for most materials J is related to the interatomic distance R through a linear relation $dJ/dR=\alpha$ (for $dR/R<<1$), we can directly derive the local distance R from the observed excitation energies. This is exemplified for the mixed one-dimensional paramagnetic compound $CsMn_xMg_{1-x}Br_3$ (x=0.05, 0.10) containing manganese dimers oriented along the hexagonal c-axis. Surprisingly, the resulting Mn-Mn distances R do not vary continuously with increasing internal pressure, but lock in at some discrete values.

Conventional crystallography is the standard tool for structure determination of materials, and a periodic lattice is a prerequisite for such studies. However, complex materials are often characterized by local deviations from perfect periodicity which may be crucial to their properties. Typical examples are the giant magnetoresistance manganates and the high-temperature superconducting cuprates which have complicated structures at the atomic or nanometer level. Different experimental techniques exist to unravel local structures, the most prominent methods being x-ray absorption fine structure (XAFS), nuclear magnetic resonance (NMR), nanobeam electron diffraction (NBED) with a transmission electron microscope (TEM) [1], and atomic pair-distribution function (PDF) analysis [2], which all yield bulk information; in addition, surface-sensitive imaging methods such as scanning tunneling microscopy (STM) and atomic force microscopy (AFM) are often used. All these methods provide a spatial resolution of typically 0.1 Å, and their performance can hardly be improved. It is therefore desirable to search for alternative techniques pushing the spatial resolution beyond the present limits. Indeed, nature is rich in relationships which connect particular physical properties of materials to their structural details. One such relationship exists for the exchange coupling J of magnetic atoms, which for most materials depends on the interatomic distance R through a linear law $dJ/dR=\alpha$ as long as $dR \ll R$. Inelastic neutron scattering experiments with the application of hydrostatic pressure performed for $Mn^{2+}$ dimers in $CsMn_{0.28}Mg_{0.78}Br_3$ [3] and $Zn_{0.98}Mn_{0.02}Te$ [4] confirmed this law with $\alpha=3.6(3)$ meV/Å and $\alpha=1.9(4)$ meV/Å, respectively. Modern spectroscopies measure exchange couplings with a precision of $dJ/J \approx 0.01$, thus spatial resolutions of $dR \approx 0.01$ Å can be achieved. This is demonstrated here for $Mn^{2+}$ dimers present in the diluted one-dimensional paramagnetic compound $CsMn_xMg_{1-x}Br_3$ (x=0.05, 0.10) whose magnetic properties are extremely well characterized [5].

$CsMn_xMg_{1-x}Br_3$ crystallizes in the hexagonal space group $P6_3/mmc$, with chains of face-sharing $MBr_6$ ($M=Mn^{2+}$, $Mg^{2+}$) octahedra along the c-axis. The partial substitution of $Mn^{2+}$ ions by non-magnetic $Mg^{2+}$ ions results in the creation of $Mn^{2+}$ monomers, dimers, trimers, etc. parallel to the c-axis as schematically sketched in Fig. 1. In the present work we focus on $Mn^{2+}$ dimer excitations whose analysis is based on the spin Hamiltonian

$$H = -2J\mathbf{s_1}\cdot\mathbf{s_2} - K(\mathbf{s_1}\cdot\mathbf{s_2})^2 - D\left[\left(s_1^z\right)^2 + \left(s_2^z\right)^2\right] \tag{1}$$

where $\mathbf{s_i}$ denotes the spin operator of the $Mn^{2+}$ ions. Within experimental error, the model parameters J, K, and D are independent of the Mn concentration x for $0.05 \leq x \leq 0.40$, with J=-0.852(3) meV and K=0.0086(2) meV being the bilinear and biquadratic exchange interactions, respectively, and D=0.0211(15) meV the axial single-ion anisotropy parameter [5]. H commutes with the square of the total spin $\mathbf{S}=\mathbf{s_1}+\mathbf{s_2}$; thus S is a good quantum number to describe the dimer spin states as |S,M> with $-S \leq M \leq S$. For antiferromagnetic exchange (J<0) the ground state is a singlet (S=0), separated from the excited triplet (S=1), quintet (S=2), etc. states according to the well known Landé interval rule E(S)-E(S-1)=-2JS. A nonzero anisotropy term (D≠0) splits the spin states |S> into the states |S,±M>. Inelastic neutron scattering experiments gave evidence for well resolved transitions out of the singlet ground state |0,0> to the excited states |1,±1> and |1,0> at energies of 1.80 meV and 1.93 meV, respectively, with an intensity ratio 2:1 [5]. The total linewidths of both transitions showed an x-dependent increase beyond the instrumental energy resolution of Gaussian shape with a full width at half maximum FWHM=55 μeV, which may be due either to further interactions in addition to those described by equation (1) or to local structural effects or to both. Therefore we investigated the higher-intensity transition in more detail with improved energy resolution.

Polycrystalline samples of $CsMn_xMg_{1-x}Br_3$ (x=0.05, 0.10) were synthesized from CsBr, $MgBr_2$, $MnBr_2$, and Mn. CsBr (Merck, suprapur) was dried at 200°C in vacuum. $MgBr_2$ was prepared from $MgCO_3 \cdot Mg(OH)_2 \cdot 3H_2O$ (Merck, pA), $NH_4Br$ (Merck, reinst), and 47% HBr acid (Merck, pA) according to the $NH_4Br$ method. $MgBr_2$ was sublimed in a gold ampoule at 640°C in vacuum. $MnBr_2$ (Cerac, 99.9%) was sublimed in a silica ampoule at 700°C in vacuum. Stoichiometric amounts of the starting materials (10 g) together with 150 mg Mn metal (Fluka, 99.9%) were sealed in silica ampoules under vacuum. The ampoules were heated to 640°C and slowly cooled to room temperature. All handling of the hygroscopic starting materials and products was done in a dry box under $N_2$ ($H_2O$ and $O_2$ both < 0.1 ppm). The samples were characterized by neutron diffraction at the spallation neutron source SINQ at PSI Villigen with use of the high-resolution diffractometer for thermal neutrons HRPT. The resulting structural parameters are listed in Table I. Within statistical error the relevant bond lengths and bond angles turned out to be independent of the dilution x.

The inelastic neutron scattering experiments were carried out using the inverted time-of-flight spectrometer MARS [6] at SINQ, which offers an unprecedented energy resolution with FWHM=15 μeV at an energy transfer of 1.8 meV. The resolution function exhibits a Lorentzian-like tail on the neutron energy-gain side as shown in Fig. 2. The neutrons scattered from the mica analyzer crystals at energy 1.87 meV were detected by $^3$He counters covering a large range of scattering angles 24°≤Θ≤156° corresponding to moduli of the scattering vector 0.60≤Q≤2.23 Å$^{-1}$. The samples were enclosed in Al cylinders (12 mm diameter, 45 mm height) and placed into a He cryostat to achieve the temperature T=1.6 K. The raw data were transformed to equidistant energy steps with use of the software package DAVE [7]. The measuring time to collect a single energy spectrum amounted to about 10 days.

The energy spectra observed for x=0.05 and x=0.10 at T=1.6 K are shown in Fig. 2. Both energy spectra exhibit marked deviations from a normal Gaussian distribution. In statistics, a criterion for relevant departures from normality is based on the sample excess kurtosis which is defined by

$$k = \frac{n^{-1}\sum_{i=1}^{n}(E_i - <E>)^4}{\left[n^{-1}\sum_{i=1}^{n}(E_i - <E>)^2\right]^2} - 3 \qquad (2)$$

where the numerator is the fourth moment about the mean energy <E> and the denominator is the square of the variance. Normal distributions have zero excess kurtosis. We applied equation (2) to the energy spectra displayed in Fig. 2 and found k=-0.96(11) for x=0.05 and k=-0.70(13) for x=0.10. Both distributions have a statistically relevant negative excess kurtosis which is called platykurtic (the Greek word πλατυσ means broad). In terms of shape, a platykurtic distribution has a wider range of data around the mean value than a normal distribution. It is therefore legitimate to analyze the data of Fig. 2 by a superposition of different lines with their lineshapes and linewidths fixed at the instrumental resolution as described above. In the least-squares analysis the fitting parameters were then a linear background as well as the number, the positions and the amplitudes of the individual lines. As shown in Fig. 2, the x=0.05 and x=0.10 data are best described by four and five individual lines, respectively. The results are listed in Table II.

What is the origin of the splitting of the |0,0>→|1,±1> transition into four (x=0.05) and five (x=0.10) individual bands? Possible mechanisms not considered in equation (1) include (i) an anisotropic exchange, (ii) a planar single-ion anisotropy, and (iii) a next-nearest-neighbor exchange coupling. The symmetry of $CsMn_xMg_{1-x}Br_3$ allows an anisotropic exchange of the form $[J(1-\varepsilon)(s^xs^x+s^ys^y)+\varepsilon s^zs^z]$ which, however, does not produce a splitting of the

|1,±1⟩ state [4], thus the mechanism (i) has to be excluded. A planar single-ion anisotropy of the form $G[(s^x)^2+(s^y)^2]$ splits the $|0,0\rangle \rightarrow |1,\pm 1\rangle$ transition into only two lines, thus the mechanism (ii) must be excluded as well. The next-nearest-neighbor exchange coupling J' was determined to be J'/J=0.014(11) [8]. $Mn^{2+}$ ions can be statistically placed on either or both sides of isolated $Mn^{2+}$ dimers as sketched in Fig. 1. For the case of a single $Mn^{2+}$ ion coupled through J' to the $Mn^{2+}$ dimer, however, we have again a splitting into only two lines. The J' coupling of two $Mn^{2+}$ ions to the $Mn^{2+}$ dimer can be discarded as well, since for our samples with x≤0.1 this local structural configuration has an extremely small probability of 1% or less, thus also the mechanism (iii) is inappropriate. More generally, the observed line splittings cannot be the result of further interactions, because the spectral shapes of the splitting patterns observed for x=0.05 and x=0.10 are quite different. We therefore conclude that the line splitting is due to structural inhomogeneities along the mixed $Mn_xMg_{1-x}$ chains, *i.e.*, each individual line of Fig. 2 corresponds to a specific $Mn^{2+}$ dimer configuration, with a particular exchange coupling $J_m$ derived from equation (1) and a particular local distance $R_m$ determined from the linear relation dJ/dR=α. Table II lists the resulting values of $J_m$ and $R_m$. The latter were derived by taking α=3.6(3) meV/Å [3], a mean exchange ⟨J⟩=-0.852(3) meV [5], and mean local distances ⟨R⟩=3.2248(1) Å (x=0.05) and ⟨R⟩=3.2256(1) Å (x=0.10) determined in the present work.

How can we explain the discrete nature of the local Mn-Mn distances $R_m$? We turn to Fig. 1 in which different configurations along the chain structure are sketched, where m is the number of $Mn^{2+}$ ions replacing the $Mg^{2+}$ ions. The introduction of additional $Mn^{2+}$ ions exerts some internal pressure within the chain, since the ionic radii of the $Mn^{2+}$ (high spin) and $Mg^{2+}$ ions are different with $r_{Mn}$=0.83 Å > $r_{Mg}$=0.72 Å [9], so that the atomic positions have to rearrange. In particular, the Mn-Mn bond distance of the central $Mn^{2+}$ dimer is expected to shorten gradually with increasing number m of $Mn^{2+}$ ions as compared to the case m=0. For any number m there is a myriad of structural configurations, resulting in a continuous distribution of local distances $R_m$. This view, however, is in contrast to the observed energy spectra displayed in Fig. 2 which are clearly not continuous but discrete with four (x=0.05) and five (x=0.10) bands. In other words, the bond distance of the central $Mn^{2+}$ dimer is not smoothly adjusted to its surrounding, but locks in at a few specific values $R_m$.

In an earlier work a model was developed to describe the structural inhomogeneities along a mixed $Mn_xMg_{1-x}$ chain [5]. Assuming a statistical distribution of $Mn^{2+}$ ions, the probabilities $p_m(x)$ for having m $Mn^{2+}$ ions on both sides of the central $Mn^{2+}$ pair in a chain of length 2n are given by

$$p_0(x) = (1-x)^{2n}$$
$$p_1(x) = 2\binom{n}{1}x(1-x)^{2n-1}$$
$$p_2(x) = \left[2\binom{n}{2} + \binom{n}{1}\binom{n}{1}\right]x^2(1-x)^{2n-2} \qquad (3)$$
$$p_3(x) = 2\left[\binom{n}{3} + \binom{n}{2}\binom{n}{1}\right]x^3(1-x)^{2n-3}$$

etc.

The chain length 2n has to be chosen such that the sum rule $\Sigma_m p_m(x)=1$ and the condition $2n \geq m$ are satisfied. In our case these criteria are fulfilled for n=1/x. Fig. 3 displays the resulting probabilities $p_m(x)$ together with the relative intensities $I_m$ of the four (x=0.05) and five (x=0.10) bands observed in the present experiments. The agreement between the experimental and the calculated data is remarkably good. Obviously the realization of discrete local distances $R_m$ is governed by the number m of $Mn^{2+}$ ions and not by their specific arrangement in the chain. At first sight this is a surprising result, since any point defect (substitutional atoms, vacancies, etc.) will create a local lattice distortion as well as displacements of neighboring atoms (located at distance r) decaying asymptotically as $1/r^2$ [10] so that one would have expected the bond shortening to be more pronounced the closer the additionally introduced $Mn^{2+}$ ions are to the central $Mn^{2+}$ dimer. The $1/r^2$ law, however, was derived for three-dimensional systems. For point defects in two-dimensional systems, the resulting displacements decrease according to $1/r$, and in one-dimensional systems the displacements do not decrease with increasing distance at all [10]. Although not strictly elastically decoupled from neighboring atoms, the linear $Mn_xMg_{1-x}$ chains may exhibit sufficient one-dimensional character to result in effective elastic distortions indifferent on the exact arrangement around the central dimer. As a net effect the mixed $Mn_xMg_{1-x}$ chains may behave like a system of hard core particles whose interaction range is not governed by short-range correlations, but extends to the nanometer scale.

The local Mn-Mn distances $R_m$ are distributed at almost equidistant steps $\Delta R_{m,m-1}=R_m-R_{m-1}\approx 0.0023$ Å. However, according to Table II and Fig. 3, the initial steps $\Delta R_{1,0}$ are slightly larger than the steps $\Delta R_{m,m-1}$ with $m\geq 2$. We attribute this observation to the statistical formation of additional $Mn^{2+}$ dimers in the chain (see Fig. 1), which are not present for m=1 but increasingly occur for $m\geq 2$. Exchange striction [3] results in a local contraction of the dimeric Mn-Mn distance by $\Delta r = 2(dJ/dR)s_1(s_1+1)/f \approx 0.012$ Å (f denotes a microscopic force constant) and thereby reduces the internal pressure within the chain. For $m\geq 2$ the probability for dimer formation is 5-10%, so that the average shrinking of the chain length amounts to 0.0006-0.0012 Å (*i.e.*, 5-10% of $\Delta r$) which is of the order of the difference between the initial steps $\Delta R_{1,0}$ and the steps $\Delta R_{m,m-1}$ with $m\geq 2$. This adds further weight to the hard core nature of the $Mn_xMg_{1-x}$ chain structure postulated above.

In conclusion, we have introduced a novel bulk tool for local structure determination based on neutron spectroscopic studies of magnetic cluster excitations. The method can be applied either to mixed compounds in which magnetic clusters are formed artificially through partial substitution by nonmagnetic ions or to pure compounds where magnetic clusters occur naturally as *e.g.* in single-molecule magnets. Our method makes specific use of the local structure dependence of the exchange interaction, but it can be extended more generally to any other property which reflects local structural effects in the atomic potentials. A promising example is provided by the crystal-field potential which is a single-ion property. Neutron crystal-field studies of the n-type high-temperature superconductor $HoBa_2Cu_3O_{7-x}$ ($0 \leq x \leq 1$) gave evidence for local displacements of the oxygen O(3) ions along the z-axis [11]. Experiments performed for the p-type high-temperature superconductor $Pr_{2-x}Ce_xCuO_4$ ($0 \leq x \leq 0.2$) with tetragonal symmetry revealed the existence of local regions whose symmetry is lower than tetragonal [12]. As long as the crystal-field potential is dominated by short-range terms, the local environments around the crystal-field active ions can be derived rather directly, similar to the method described in the present work. This is in contrast to other techniques like XAFS, NBED/TEM, and PDF analyses which usually have to be combined with simulations.

**Acknowledgments**

This work was performed at the Swiss Spallation Neutron Source SINQ, Paul Scherrer Institut (PSI), Villigen, Switzerland. Financial support by the Swiss National Science Foundation through the NCCR MaNEP project is gratefully acknowledged. We wish to dedicate this work to our colleague Philip Tregenna-Piggott who participated in the early stages of the experiments but unexpectedly passed away last year.

TABLE I. Crystal structure parameters, selected bond distances d and bond angles φ determined for $CsMn_xMg_{1-x}Br_3$ (x=0.05, 0.10) in the structure model $P6_3/mmc$ (No. 194). Mn and Mg are in the (2a)-position (0,0,0), Cs in (2d) (1/3, 2/3, 3/4), and Br in (12j) (x,y, 1/4).

|  | x=0.05 | | x=0.10 | |
| --- | --- | --- | --- | --- |
|  | T=1.5 K | T=300 K | T=1.5 K | T=300 K |
| a [Å] | 7.53192(5) | 7.60968(7) | 7.53255(6) | 7.61082(8) |
| c [Å] | 6.44962(6) | 6.50476(9) | 6.45121(7) | 6.50648(9) |
| x | 0.16149(22) | 0.16055(24) | 0.16138(25) | 0.16008(26) |
| y | 0.32299(45) | 0.32111(49) | 0.32277(51) | 0.32017(52) |
| $B_{Mn,Mg}$ [Å²] | 0.706(53) | 1.609(60) | 0.353(69) | 1.420(71) |
| $B_{Cs}$ [Å²] | 0.227(48) | 3.053(84) | 0.469(57) | 2.919(89) |
| $B_{Br}$ [Å²] | 0.241(18) | 1.926(22) | 0.350(21) | 1.912(25) |
| $\chi^2$ | 5.86 | 3.49 | 9.34 | 4.20 |
| $d_{Mn-Br}$ [Å] | 2.6531(15) | 2.6688(15) | 2.652(2) | 2.665(2) |
| $\varphi_{Mn-Br-Mn}$ [°] | 74.85(3) | 75.08(3) | 74.90(5) | 75.25(5) |

TABLE II. Summary and analysis of experimental data observed for $CsMn_xMg_{1-x}Br_3$. $E_m$ and $I_m$ denote the energy transfers and the normalized intensities, respectively, of the individual lines displayed in Fig. 2. The exchange couplings $J_m$ and the local Mn-Mn distances $R_m$ associated with each line m were derived as explained in the text. For both $J_m$ and $R_m$ relative error bars are given.

| x    | m | $E_m$ [meV] | $I_m$      | $J_m$ [meV]  | $R_m$ [Å]   |
|------|---|-------------|------------|--------------|-------------|
| 0.05 | 0 | 1.762(4)    | 0.115(37)  | -0.8321(19)  | 3.2303(6)   |
| 0.05 | 1 | 1.782(2)    | 0.418(47)  | -0.8420(10)  | 3.2276(3)   |
| 0.05 | 2 | 1.796(3)    | 0.343(46)  | -0.8488(15)  | 3.2257(4)   |
| 0.05 | 3 | 1.813(4)    | 0.124(39)  | -0.8573(19)  | 3.2233(6)   |
| 0.10 | 0 | 1.762(2)    | 0.081(22)  | -0.8321(10)  | 3.2311(3)   |
| 0.10 | 1 | 1.780(2)    | 0.302(25)  | -0.8408(10)  | 3.2287(3)   |
| 0.10 | 2 | 1.794(2)    | 0.383(22)  | -0.8477(10)  | 3.2268(3)   |
| 0.10 | 3 | 1.811(2)    | 0.173(22)  | -0.8563(10)  | 3.2244(3)   |
| 0.10 | 4 | 1.826(3)    | 0.061(18)  | -0.8637(14)  | 3.2224(4)   |

**Figure Captions**

FIG. 1. (Color online) Sketch of statistical distributions of $Mn^{2+}$ and $Mg^{2+}$ ions in $CsMn_xMg_{1-x}Br_3$ along the c-axis. m is the number of $Mn^{2+}$ ions replacing $Mg^{2+}$ ions. 2n denotes the chain length outside the central $Mn^{2+}$ dimer embedded in the brown area, and m is the number of peripheral $Mn^{2+}$ ions in the chain. The blue area marks the case of a single $Mn^{2+}$ ion coupled through the next-nearest-neighbor exchange J' to the $Mn^{2+}$ dimer, thereby forming a J-J' trimer. The red area indicates the J' coupling of two $Mn^{2+}$ ions to the $Mn^{2+}$ dimer, resulting in a J'-J-J' tetramer. The green areas mark the statistical formation of additional $Mn^{2+}$ dimers in the chain.

FIG. 2. (Color online) Energy distributions of the $|0,0\rangle \rightarrow |1,\pm 1\rangle$ $Mn^{2+}$ dimer transition observed for $CsMn_xMg_{1-x}Br_3$ (x=0.05, 0.10) at T=1.6 K. The full lines are the result of a least-squares fitting procedure as explained in the text. The dashed lines show the subdivision into individual bands.

FIG. 3. (Color online) Distribution of local Mn-Mn distances asociated with $Mn^{2+}$ dimers in $CsMn_xMg_{1-x}Br_3$ (x=0.05, 0.10). The experimental data (fractional intensities $I_m$ taken from Table II) are compared to the probabilities $p_m(x)$ predicted by the statistical model according to equation (3). The widths of the intensity profiles correspond to the error bars of $R_m$ listed in Table II.

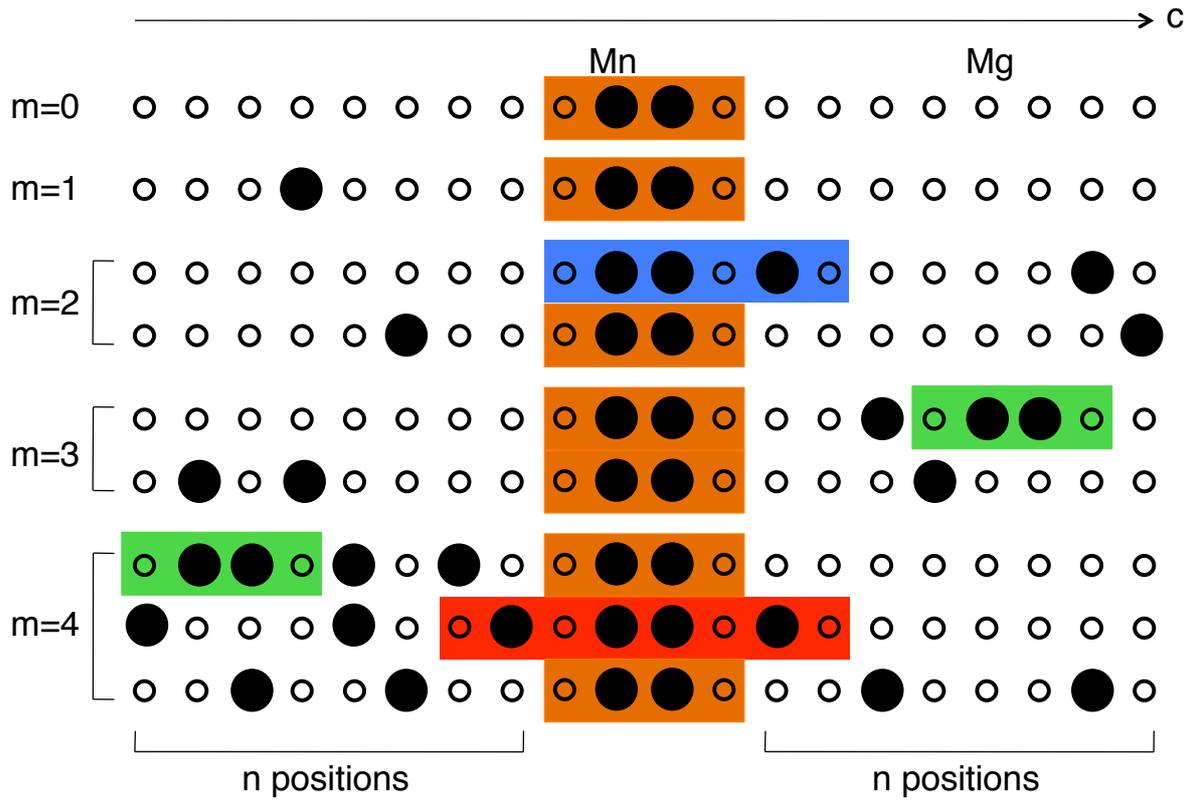

FIG. 1. (Color online) Sketch of statistical distributions of $Mn^{2+}$ and $Mg^{2+}$ ions in $CsMn_xMg_{1-x}Br_3$ along the c-axis. m is the number of $Mn^{2+}$ ions replacing $Mg^{2+}$ ions. 2n denotes the chain length outside the central $Mn^{2+}$ dimer embedded in the brown area, and m is the number of peripheral $Mn^{2+}$ ions in the chain. The blue area marks the case of a single $Mn^{2+}$ ion coupled through the next-nearest-neighbor exchange J' to the $Mn^{2+}$ dimer, thereby forming a J-J' trimer. The red area indicates the J' coupling of two $Mn^{2+}$ ions to the $Mn^{2+}$ dimer, resulting in a J'-J-J' tetramer. The green areas mark the statistical formation of additional $Mn^{2+}$ dimers in the chain.

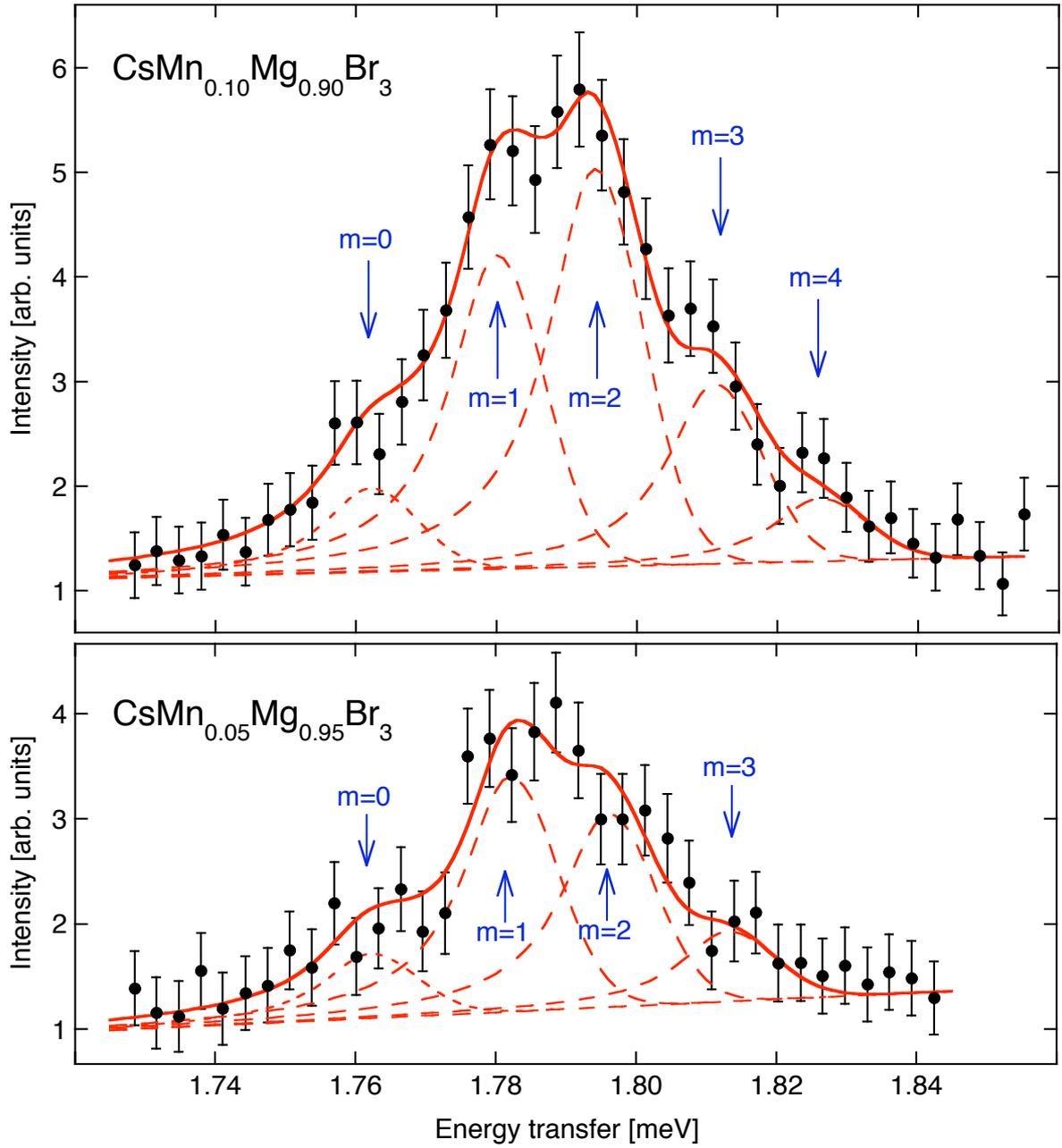

FIG. 2. (Color online) Energy distributions of the $|0,0\rangle \rightarrow |1,\pm 1\rangle$ $Mn^{2+}$ dimer transition observed for $CsMn_xMg_{1-x}Br_3$ (x=0.05, 0.10) at T=1.6 K. The full lines are the result of a least-squares fitting procedure as explained in the text. The dashed lines show the subdivision into individual bands.

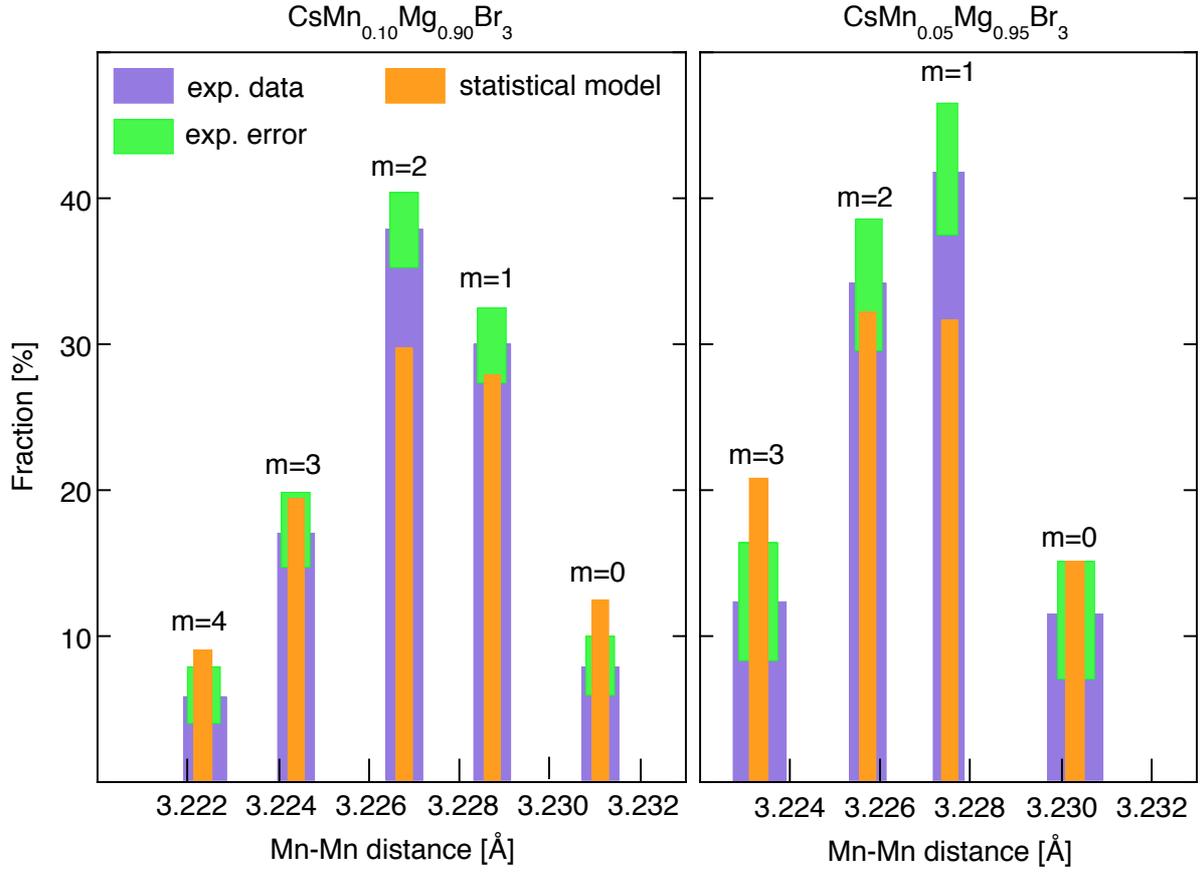

FIG. 3. (Color online) Distribution of local Mn-Mn distances asociated with $Mn^{2+}$ dimers in $CsMn_xMg_{1-x}Br_3$ (x=0.05, 0.10). The experimental data (fractional intensities $I_m$ taken from Table II) are compared to the probabilities $p_m(x)$ predicted by the statistical model according to equation (3). The widths of the intensity profiles correspond to the error bars of $R_m$ listed in Table II.